\documentclass[aps,prl,twocolumn,superscriptaddress,showpacs,floatfix]{revtex4}

\usepackage[dvips]{graphicx}
\usepackage{xspace}
\usepackage{latexsym}
\usepackage{amssymb}
\usepackage{dcolumn}
\usepackage{bm}
\usepackage{times}

\begin{document}

\newcommand{\superk}    {Super-K\xspace}
\newcommand{\nue}       {$\nu_{e}$\xspace}
\newcommand{\numu}      {$\nu_{\mu}$\xspace}
\newcommand{\nutau}     {$\nu_{\tau}$\xspace}
\newcommand{\nusterile} {$\nu_{sterile}$\xspace}
\newcommand{\mue}       {$\nu_\mu \leftrightarrow \nu_{e}$\xspace}
\newcommand{\mutau}     {$\nu_\mu \leftrightarrow \nu_{\tau}$\xspace}
\newcommand{\musterile} {$\nu_\mu \leftrightarrow \nu_{sterile}$\xspace}
\newcommand{\dms}       {$\Delta m^2$\xspace}
\newcommand{\sstt}      {$\sin^2 2 \theta$\xspace}

\title{A Measurement of Atmospheric Neutrino Flux Consistent with Tau Neutrino Appearance}

\date{\today}

\newcommand{\icrr}{\affiliation{Kamioka Observatory, Institute for Cosmic Ray Research, University of Tokyo, Kamioka, Gifu, 506-1205, Japan}}
\newcommand{\ncen}{\affiliation{Research Center for Cosmic Neutrinos, Institute for Cosmic Ray Research, University of Tokyo, Kashiwa, Chiba 277-8582, Japan}}
\newcommand{\bu}{\affiliation{Department of Physics, Boston University, Boston, MA 02215, USA}}
\newcommand{\bnl}{\affiliation{Physics Department, Brookhaven National Laboratory, Upton, NY 11973, USA}}
\newcommand{\uci}{\affiliation{Department of Physics and Astronomy, University of California, Irvine, Irvine, CA 92697-4575, USA}}
\newcommand{\csu}{\affiliation{Department of Physics, California State University, Dominguez Hills, Carson, CA 90747, USA}}
\newcommand{\cnu}{\affiliation{Department of Physics, Chonnam National University, Kwangju 500-757, Korea}}
\newcommand{\duke}{\affiliation{Department of Physics, Duke University, Durham, NC 27708 USA}}
\newcommand{\gmu}{\affiliation{Department of Physics, George Mason University, Fairfax, VA 22030, USA}}
\newcommand{\gifu}{\affiliation{Department of Physics, Gifu University, Gifu, Gifu 501-1193, Japan}}
\newcommand{\uh}{\affiliation{Department of Physics and Astronomy, University of Hawaii, Honolulu, HI 96822, USA}}
\newcommand{\ui}{\affiliation{Department of Physics, Indiana University, Bloomington,  IN 47405-7105, USA} }
\newcommand{\kek}{\affiliation{High Energy Accelerator Research Organization (KEK), Tsukuba, Ibaraki 305-0801, Japan}}
\newcommand{\kobe}{\affiliation{Department of Physics, Kobe University, Kobe, Hyogo 657-8501, Japan}}
\newcommand{\kyoto}{\affiliation{Department of Physics, Kyoto University, Kyoto 606-8502, Japan}}
\newcommand{\lanl}{\affiliation{Physics Division, P-23, Los Alamos National Laboratory, Los Alamos, NM 87544, USA}}
\newcommand{\lsu}{\affiliation{Department of Physics and Astronomy, Louisiana State University, Baton Rouge, LA 70803, USA}}
\newcommand{\umd}{\affiliation{Department of Physics, University of Maryland, College Park, MD 20742, USA}}
\newcommand{\MIT}{\affiliation{Department of Physics, Massachusetts Institute of Technology, Cambridge, MA 02139, USA}}
\newcommand{\duluth}{\affiliation{Department of Physics, University of Minnesota, Duluth, MN 55812-2496, USA}}
\newcommand{\miyagi}{\affiliation{Department of Physics, Miyagi University of Education, Sendai,Miyagi 980-0845, Japan}}
\newcommand{\suny}{\affiliation{Department of Physics and Astronomy, State University of New York, Stony Brook, NY 11794-3800, USA}}
\newcommand{\nagoya}{\affiliation{Department of Physics, Nagoya University, Nagoya, Aichi 464-8602, Japan}}
\newcommand{\nagoyaste}{\affiliation{Solar-Terrestrial Environment Laboratory, Nagoya University, Nagoya, Aichi 464-8601, Japan}}
\newcommand{\niigata}{\affiliation{Department of Physics, Niigata University, Niigata, Niigata 950-2181, Japan}}
\newcommand{\osaka}{\affiliation{Department of Physics, Osaka University, Toyonaka, Osaka 560-0043, Japan}}
\newcommand{\okayama}{\affiliation{Department of Physics, Okayama University, Okayama, Okayama 700-8530, Japan}}
\newcommand{\seoul}{\affiliation{Department of Physics, Seoul National University, Seoul 151-742, Korea}}
\newcommand{\shizuokaseika}{\affiliation{International and Cultural Studies, Shizuoka Seika College, Yaizu, Shizuoka 425-8611, Japan}}
\newcommand{\shizuokafukushi}{\affiliation{Department of Informatics in
Social Welfare, Shizuoka University of Welfare, Yaizu, Shizuoka 425-8611, Japan}}
\newcommand{\shizuoka}{\affiliation{Department of Systems Engineering, Shizuoka University, Hamamatsu, Shizuoka 432-8561, Japan}}
\newcommand{\skku}{\affiliation{Department of Physics, Sungkyunkwan University, Suwon 440-746, Korea}}
\newcommand{\tohoku}{\affiliation{Research Center for Neutrino Science, Tohoku University, Sendai, Miyagi 980-8578, Japan}}
\newcommand{\tokyo}{\affiliation{University of Tokyo, Tokyo 113-0033, Japan}}
\newcommand{\tokai}{\affiliation{Department of Physics, Tokai University, Hiratsuka, Kanagawa 259-1292, Japan}}
\newcommand{\tit}{\affiliation{Department of Physics, Tokyo Institute for Technology, Meguro, Tokyo 152-8551, Japan}}
\newcommand{\warsaw}{\affiliation{Institute of Experimental Physics, Warsaw University, 00-681 Warsaw, Poland}}
\newcommand{\uw}{\affiliation{Department of Physics, University of Washington, Seattle, WA 98195-1560, USA}}
\newcommand{\tsukubanow}{\altaffiliation{ Present address: Department of Physics, Univ. of Tsukuba, Tsukuba, Ibaraki 305 8577, Japan}}
\newcommand{\okayamanow}{\altaffiliation{ Present address: Department of Physics, Okayama University, Okayama 700-8530, Japan}}
\newcommand{\marylandnow}{\altaffiliation{ Present address: University of Maryland School of Medicine, Baltimore, MD 21201, USA}}
\newcommand{\triunfnow}{\altaffiliation{ Present address: TRIUMF, Vancouver, British Columbia V6T 2A3, Canada}}
\newcommand{\icrrnow}{\altaffiliation{ Present address: Kamioka Observatory, Institute for Cosmic Ray Research, University of Tokyo, Kamioka, Gifu, 506-1205, Japan}}
\newcommand{\pennnow}{\altaffiliation{ Present address: Center for Gravitational Wave Physics, Pennsylvania State University, University Park, PA 16802, USA}}
%
\author{K.Abe}\icrr
\author{Y.Hayato}\icrr
\author{T.Iida}\icrr
\author{K.Ishihara}\icrr
\author{J.Kameda}\icrr
\author{Y.Koshio}\icrr
\author{A.Minamino}\icrr
\author{C.Mitsuda}\icrr
\author{M.Miura}\icrr
\author{S.Moriyama}\icrr
% ...
\collaboration{The Super-Kamiokande Collaboration}\noaffiliation

\begin{abstract}
A search for the appearance of tau neutrinos from \mutau oscillations
in the atmospheric neutrinos has been performed using 1489.2 days of 
atmospheric neutrino data from the Super-Kamiokande-I experiment.
A best fit tau neutrino appearance signal of 
138\,$\pm$\,48\,(stat.)\,$^{+15}_{-32}$\,(sys.) events
is obtained with an expectation of 78\,$\pm$\,26\,(sys.). 
The hypothesis of no tau neutrino appearance is disfavored by 2.4 sigma.
\end{abstract}

\pacs{14.60.Pq, 95.55.Vj, 95.85.Ry} 


\maketitle

Atmospheric neutrino oscillations have been observed by several
experiments~\cite{Fukuda:1998mi, Fukuda:1994mc, 
Ambrosio:2004ig, Allison:2005dt,  Adamson:2005qc}; 
in particular, Super-Kamiokande (Super-K) 
has reported the first evidence for the sinusoidal signature of muon neutrino
disappearance~\cite{Ashie:2004mr} and made measurements of the
oscillation parameters~\cite{Ashie:2005ik}. The {\superk} atmospheric
neutrino data favor \mutau oscillations
and have excluded \mue~\cite{Fukuda:1998mi} and 
pure {\musterile} oscillations~\cite{Fukuda:2000np} 
as a dominant source of the deficit of muon neutrinos.
All of these results rely largely on the disappearance of 
muon neutrinos from the atmospheric neutrino flux, with no explicit 
observation of the appearance of tau neutrinos 
and their charged-current (CC) weak interactions. 
In this letter, we analyze the Super-K atmospheric 
neutrino data in an attempt to demonstrate the appearance of \nutau 
interactions in the detector.

Detection of CC \nutau interactions in the atmospheric 
neutrinos is challenging for two reasons. First, the neutrino energy
threshold for tau lepton production is 3.5\,GeV. The atmospheric
neutrino flux above this energy is relatively low. Assuming two flavor
maximal mixing of \mutau with \dms = $2.4 \times 10^{-3}$\,eV$^2$, 
approximately one CC \nutau event is expected to
occur in an atmospheric neutrino detector per kiloton-year of exposure. 
This corresponds to 
an estimated total of 78 \nutau events in the data sample presented. 
Second, the tau lepton 
has a short lifetime (290 femtoseconds) and decays immediately into 
many different final states. These final states consist of electrons, muons, 
or one or more pions (plus always a tau neutrino). The recoiling hadronic 
system may also produce multiple particles. Water Cherenkov detectors such as
\superk are not suited for identifying individual CC \nutau interactions
as there are generally multiple Cherenkov rings with no easily
identified leading lepton. Thus, we employ likelihood and neural
network techniques to discriminate tau neutrino events from
atmospheric neutrino events on a statistical basis. 

Super-Kamiokande is a 50-kton water Cherenkov detector, with a rock
overburden of 2700\,m water equivalent, located in Kamioka, Gifu prefecture 
in Japan. 
The detector consists of two concentric optically separated detector
regions; the inner detector (ID) instrumented with 11,146 inward 
facing 20\,inch diameter photomultiplier tubes (PMT) and 
the outer detector (OD) instrumented with 1,885 outward facing 8\,inch PMTs.
The details of the detector, calibrations, data reduction, and
detector simulation can be found in Refs.~\cite{Ashie:2005ik,Fukuda:2002uc}.

In this letter, the atmospheric neutrino data accumulated during the \superk-I 
period (from May, 1996 to July, 2001; 1489.2 live-day exposure) 
are analyzed. 
The atmospheric neutrino events in \superk are classified as
fully-contained (FC), partially-contained (PC), and upward-going muon
events~\cite{Ashie:2005ik}.  In the present
analysis, only FC events are used.  FC events deposit all of their
Cherenkov light inside the ID, from which the direction and the momentum 
of charged particles are reconstructed.
The particle type is identified as ``$e$-like (showering)'' 
or ``$\mu$-like (non-showering)'' for each Cherenkov ring 
based on the light-pattern~\cite{Ashie:2005ik}.

Both tau neutrino and atmospheric neutrino (\nue and \numu) 
interactions in \superk are modeled using a Monte Carlo (MC)
simulation, NEUT~\cite{Ashie:2005ik, Hayato:2002sd}.  
In our \nutau MC, only CC \nutau interactions are simulated. 
CC \nutau interactions are mostly deep-inelastic scattering 
(approximately 63\%) due to the high energy threshold for 
tau lepton production. 
The decays of tau leptons are simulated by the tau decay library, 
TAUOLA (Version 2.6)~\cite{Jadach:1993hs}. The polarization of tau leptons 
produced via CC \nutau interactions is also implemented from calculations by 
Hagiwara {\it et al.}~\cite{Hagiwara:2003di}.

The event signatures of tau neutrinos are characterized by the decays of 
tau leptons produced in CC \nutau interactions. In this analysis, 
we concentrate on the hadronic decays of tau leptons 
(approximately 65\% of tau lepton decays). 
The shape of events containing the hadronic decays of tau leptons has a more
spherical topology than that of backgrounds. Also, the extra pions produced 
in tau lepton decays can be tagged by looking for their decays and 
ring signatures in the \superk detector. 
The primary backgrounds for \nutau signals are atmospheric neutrinos 
producing multiple pions via deep-inelastic scattering interactions.
The following \nutau event selection criteria are applied 
to reduce the backgrounds: 
(1) The vertex must be reconstructed 
in the fiducial volume (2\,m from the ID PMT surface) with no activity 
in the OD region,
(2) Visible energy ($E_{vis}$) must be greater than 1.33\,GeV, and 
(3) The most energetic ring must be $e$-like.  
These criteria can reduce the backgrounds by approximately 90\% 
since the neutrino energy threshold in CC \nutau interactions is higher than 
in most of atmospheric neutrino interactions and 
tau lepton decays have a high average multiplicity (resulting primarily in 
hadronic shower, i.e. $e$-like events). 
The effects of the event selection criteria 
are summarized in the top 3 lines of Table~\ref{table:event}.
\begin{table}[bp]
  \begin{tabular}{lccc}
    \hline
    \hline
                       & ~Data~ & ~$\nu_{e,\mu}$ BKG MC~ & ~CC \nutau MC~ \\
    \hline
    Generated in fiducial volume&  --  & 17135(100\%)  & 78.4 (100\%) \\
    E$_{vis}$ $>$ 1.33 GeV      & 2888 & 2943 (17.2\%) & 51.5 (65.7\%)\\
    Most energetic ring e-like  & 1803 & 1765 (10.3\%) & 47.1 (60.1\%)\\
    \hline
    Likelihood $>$ 0            & 649  & 647 (3.8\%)   & 33.8 (43.1\%)\\
    \hline
    NN output $>$ 0.5           & 603  & 577 (3.4\%)   & 30.6 (39.0\%)\\
    \hline
    \hline
  \end{tabular}
  \caption{Summary for the numbers of events in data, atmospheric neutrino ($\nu_{e},_{\mu}$) background MC, and tau neutrino MC after applying each of \nutau event selection criteria and either the likelihood or the neural network cut, which is applied separately for each analysis. Neutrino oscillation is considered in the MCs with the oscillation parameters: \dms = $2.4 \times 10^{-3}$\,eV$^2$ and \sstt = 1.0, and the numbers are normalized by the live time of the data sample.}
  \label{table:event}
\end{table}

\begin{figure}[tbp]
  \includegraphics[width=3.1in]{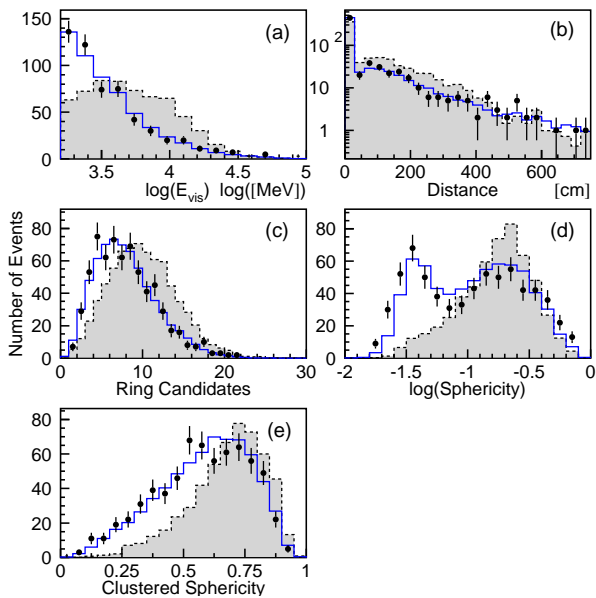}
  \caption{The distributions of variables:
    (a) Visible energy,
    (b) Maximum distance between the primary interaction and electron vertices from pions and then $\mu$ decays, 
    (c) Number of ring candidates,
    (d) Sphericity in the laboratory frame, and
    (e) Clustered sphericity in the center of mass, 
    after applying the \nutau event selection criteria 
    for downward-going data (points),
    \nutau MC (shaded histogram), 
    and atmospheric $\nu_{e},_{\mu}$ background MC events (solid histogram). 
    (The histograms of \nutau MC are normalized arbitrarily.) 
    In the likelihood analysis, the data sample is divided into 5 energy bins.}
  \label{figure:variables}
\end{figure}

In accordance with the event shape and characteristics of tau lepton decays,
we define a set of five variables 
to further discriminate \nutau signals from backgrounds, which are: 
(a) Visible energy, 
(b) Maximum distance between the primary interaction 
and electron vertices from pion and then muon decays, 
(c) Number of ring candidates, 
(d) Sphericity in the laboratory frame, and 
(e) Clustered sphericity in the center of mass frame 
(FIG.~\ref{figure:variables}).
The first two variables are \superk standard variables~\cite{Ashie:2005ik}. 
The number of ring candidates is determined with a ring-finding algorithm, 
which is sensitive to ring-fragments, and the last two variables, sphericity 
and clustered sphericity, are derived from a jet-based energy flow analysis.
 
The energy flow analysis uses Cherenkov patterns to deconvolve the
measured light distribution in \superk.  By associating the deconvoluted
power spectrum with pseudo particles, "jets" are reconstructed, and event
shape variables such as sphericity are obtained.
Sphericity (0 $<$ S $<$ 1) measures the spherical symmetry of an event
and has a quadratic momentum dependence giving more weight to higher momentum 
particles~\cite{Bjorken:1969wi,Hanson:1975fe}. Also, it is not invariant under 
the Lorentz transformation. Therefore, the sphericity calculated with 
different clustering and reference frames probes different event 
characteristics.

The expected distributions of each variable after applying the \nutau 
event selection criteria for both signals and backgrounds 
are plotted in FIG.~\ref{figure:variables}.
The background MC is compared with downward-going data, where no 
\nutau appearance signal is expected as the probability for \numu
oscillating into \nutau is very small for the given path length and 
the measured atmospheric \dms.
The agreement of the downward-going data and background MC indicates the
variables chosen for this analysis are well modeled by our MC simulation.
The small discrepancy in log(Sphericity) between the data and the atmospheric 
neutrino MC is consistent with the systematic uncertainty 
in the vertex position.

\begin{figure}[tbp]
  \includegraphics[width=3.3in]{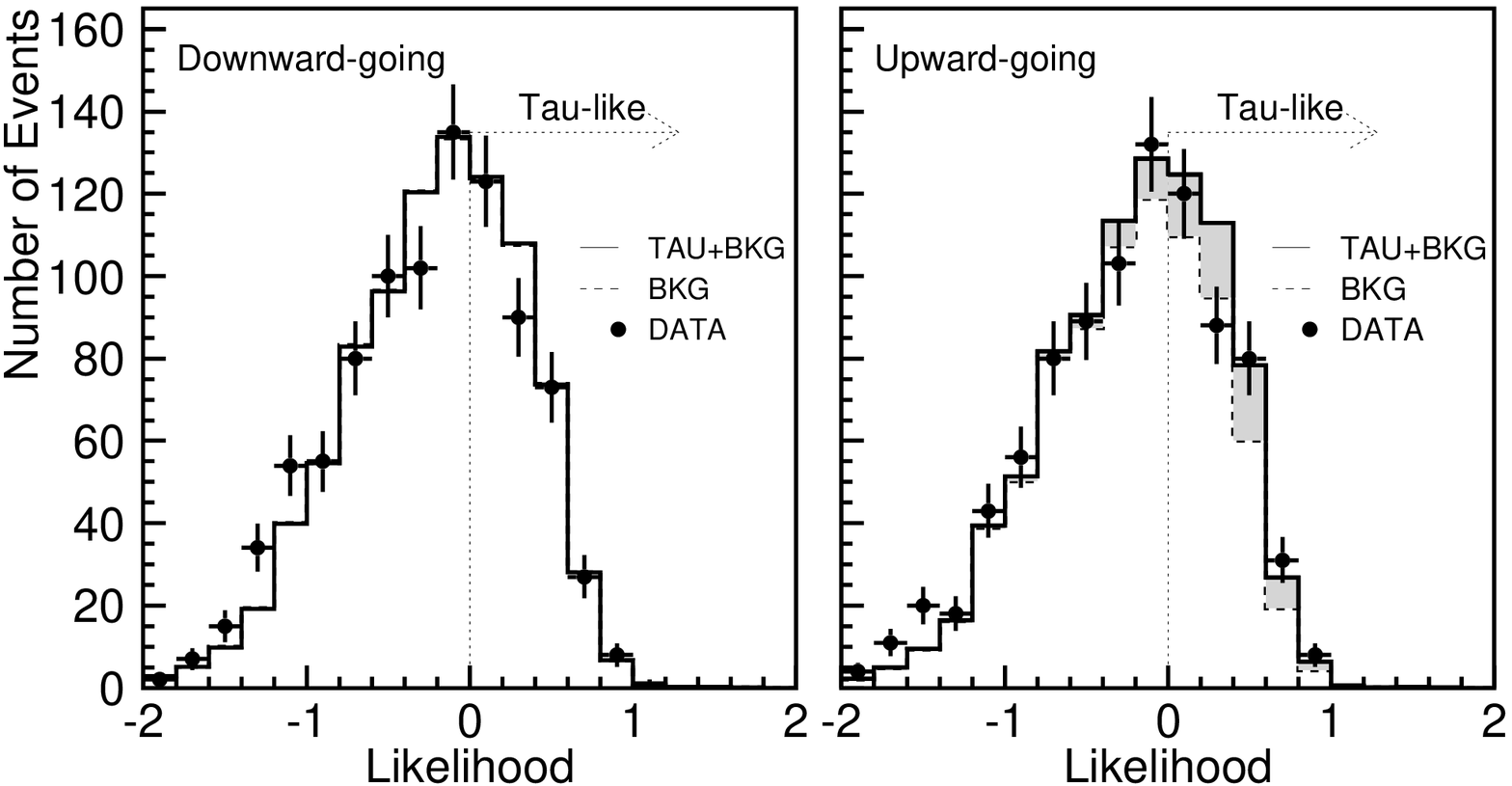}
  \includegraphics[width=3.3in]{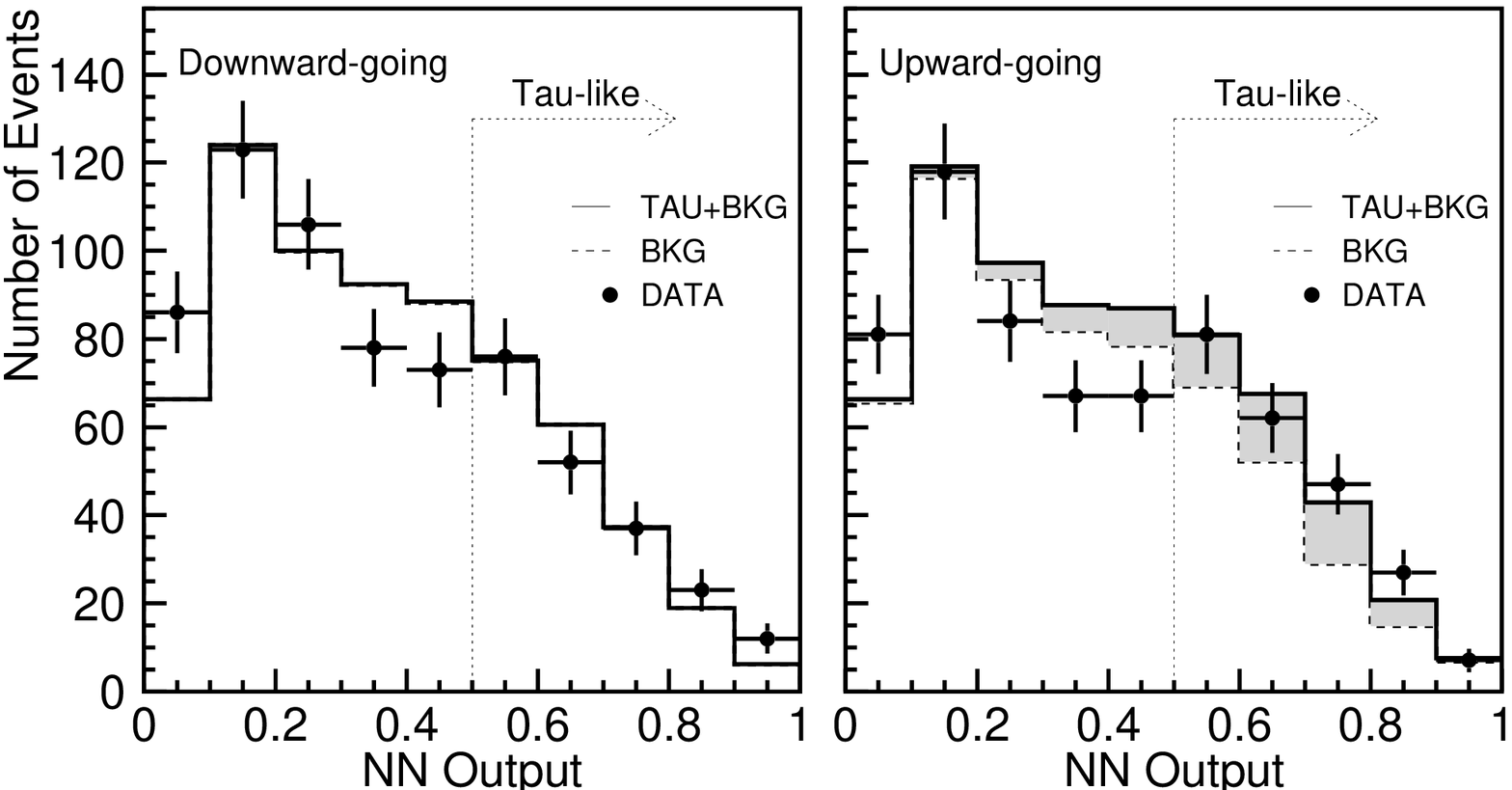}
  \caption{The likelihood (top) and NN output (bottom) distributions of downward-going (left) and upward-going (right) events for data (points), atmospheric neutrino background MC (dashed histogram), and the best fit including tau neutrino and backgrounds (solid histogram). The shaded area shows a fitted excess of tau neutrino events in the upward-going direction. The events for likelihood $\cal L$ $>$ 0 or NN output $>$ 0.5 are defined to be tau-like.}
  \label{figure:likelihood}
\end{figure}

We have constructed a likelihood function using the five variables 
described above. 
The data sample is divided into 5 energy bins: 
(1) $ E_{vis} < $ 2.0, (2) 2.0 $\leq E_{vis} < $ 3.0, 
(3) 3.0 $\leq E_{vis} < $ 6.0, (4) 6.0 $\leq E_{vis} < $ 12.0, 
(5) 12.0 $\leq E_{vis}$ [GeV].
The likelihood distributions for downward-going and upward-going events 
are shown in FIG.~\ref{figure:likelihood}. 
The events for likelihood $\cal L >$ 0 are defined to be tau-like. 
The likelihood distributions of data and background MC events agree 
for downward-going events.
The agreement validates our analysis method.
Table~\ref{table:event} summarizes the number of data, atmospheric neutrino 
background MC, and tau neutrino MC events after applying 
all of the event selection criteria.

A neural network (NN) is also trained with the five variables. 
The network has 6 input neurons, 10 hidden neurons, and
one sigmoid output neuron and is trained using back-propagation by
use of the MLPFIT neural network package~\cite{MLP-ref}.  

The distributions of the NN outputs are shown in FIG.~\ref{figure:likelihood}. 
NN output $>$ 0.5 is defined to be tau-like. 
The discrepancy between the data and MC just below the NN cut 
(NN output = 0.5) is consistent with the systematic uncertainty 
in the deep-inelastic scattering cross section near the threshold, 
which is included in the overall systematic uncertainty estimation 
described below. 

\begin{figure}[htbp]
  \includegraphics[width=2.5in]{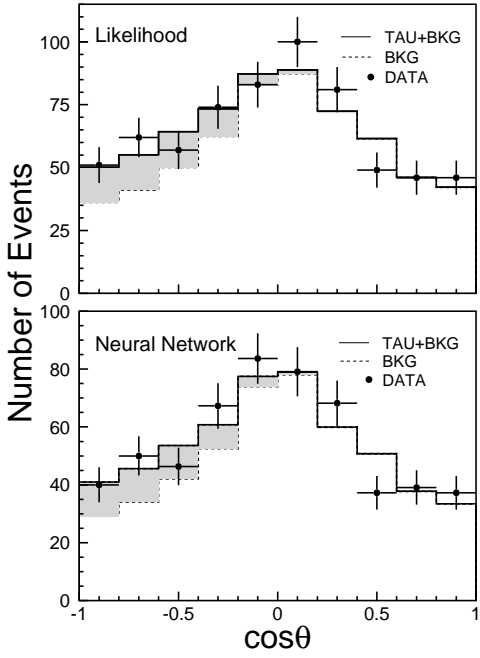}
  \caption{The zenith angle distributions for the likelihood (top) and neural network (bottom) analyses. Zenith angle cos$\theta$ = $-$1 (cos$\theta$ = $+$1) indicates upward-going (downward-going). The data sample is fitted after \nutau event selection criteria are applied. The solid histogram shows the best fit including \nutau, and the dashed histogram shows the backgrounds from atmospheric neutrinos (\nue and \numu). A fitted excess of tau-like events in the upward-going direction is shown in the shaded area.}
   \label{figure:zenith_fit}
\end{figure}
After selecting the tau-enriched sample by 
applying the \nutau event selection criteria with either the likelihood 
($\cal L >$ 0) or the neural network (NN output $>$ 0.5) cut, 
the zenith angle distribution is fitted with a combination of 
the expected tau neutrino signals resulting from oscillations 
and the predicted atmospheric neutrino background events including 
oscillations. The fitted zenith angle distribution and the $\chi^2$, 
which is minimized, are
\begin{eqnarray}
  N_{\rm total}( \cos\theta ) = \alpha N_{\rm tau} + \beta N_{\rm bkg}, \\
  \label{zenith}
  \chi^2 =  \sum_{i=1}^{10}
  \frac{\left(N_{i}^{\rm obs} -\alpha N_{i}^{\rm tau} - \beta N_{i}^{\rm bkg}\right)^2}{ \sigma^2_{i} },
  \label{chi2}
\end{eqnarray}
\noindent
where $N^{\rm obs}_i$ is the number of the observed events, 
$N^{\rm tau}_i$ is the number of predicted tau neutrino events, 
$N^{\rm bkg}_i$ is the MC predicted number of atmospheric neutrino
background events, and $\sigma_{i}$ is the statistical error 
for the $i$-th bin. The sample normalizations, $\alpha$ and $\beta$, are 
allowed to vary freely, and the zenith angle distribution is divided 
into 10 bins, from $-$1 to 1 (cos$\theta$ = $-$1 (cos$\theta$ = 1) 
refers to upward-going (downward-going) events). 
The results of the fit are shown in FIG.~\ref{figure:zenith_fit}.

The minimum $\chi^2$ for the likelihood fit (the NN fit) is 
$\chi^2_{min} = 7.6/8 {\rm ~DOF}$ ($9.8/8 {\rm ~DOF}$),
and the $\chi^2$ assuming no tau neutrino appearance is $16.3/9 {\rm ~DOF}$ 
($18.2/9 {\rm ~DOF}$).
The normalizations of the best fit for the likelihood fit (the NN fit) 
are $\alpha$ = 1.76 (1.71) and $\beta$ = 0.90 (0.99). 
After correcting for efficiencies, these correspond to a best fit tau neutrino 
appearance signal of 138\,$\pm$\,48\,(stat.) (134\,$\pm$\,48\,(stat.)) 
for the likelihood analysis (the NN analysis).  As can be seen in
FIG.~\ref{figure:zenith_fit}, an excess of tau neutrino signals is observed
in the upward-going direction, and the data distribution agrees better
with the prediction including tau neutrino appearance estimated by MC.  
The backgrounds that remain after applying all of the \nutau
event selection criteria are mostly deep-inelastic scattering (CC DIS: 61.4\% 
and NC DIS: 27.1\%).

Approximately 82.9\,$\pm$\,3.0\,\% of events are in common to 
the tau-enriched samples selected by both analyses, for which 
Monte Carlo predicts 83.1\% of events overlap.
The results for the likelihood and the neural network analyses 
are consistent.

\begin{table}[tbp]
    \begin{tabular}{lrr}
      \hline 
      \hline
      Systematic uncertainties for expected \nutau & ~~LH (\%) & ~~~~NN (\%)\\
      \hline
      \superk atmospheric $\nu$ oscillation analysis & 21.6 & 20.2 ~~\\ 
      (23 error terms)  &   &  \\
      \hline
      Tau related:      &   & \\
      {\hspace{1cm}}Tau neutrino cross section         & 25.0 & 25.0 ~~\\
      {\hspace{1cm}}Tau lepton polarization            & 7.2  & 11.8 ~~\\
      {\hspace{1cm}}Tau neutrino selection efficiency  & 0.4  & 0.5  ~~\\
      {\hspace{1cm}}LH selection efficiency        & 4.8   & --~~  ~~\\
      {\hspace{1cm}}NN selection efficiency        &  --~~ & 3.0   ~~\\
      \hline
      Total:  &  32.6 & 34.4 ~~\\
      \hline
      \hline
    \end{tabular}

    {\vspace{0.2cm}}

    \begin{tabular}{lrr}
      \hline
      \hline
      Systematic uncertainties for observed \nutau & ~~LH (\%) & ~~~NN (\%)\\
      \hline
      \superk atmospheric $\nu$ oscillation analysis: & & \\
      {\hspace{1cm}}Flux up/down ratio & 6.5 & 5.7 ~~\\
      {\hspace{1cm}}Flux horizontal/vertical ratio & 3.6 & 3.2 ~~\\
      {\hspace{1cm}}Flux K/$\pi$ ratio & 2.4 & 2.8 ~~\\
      {\hspace{1cm}}NC/CC ratio & 4.3 & 3.8 ~~\\
      {\hspace{1cm}}Up/down asym. from energy calib. & 1.4 & $<$ 0.1 ~~\\
      Oscillation parameters:   &    & \\
      {\hspace{1cm}}$0.0020 < \Delta m^2_{23} < 0.0027 \,{\rm eV}^2$ & $+$5.8 & $+$8.8 ~~\\
                                  & $-$2.6 & $-$3.3 ~~\\
      {\hspace{1cm}}$0.93 < \sin^{2}2\theta_{23} < 1.00$ & $-$3.3 & $-$3.9 ~~\\
      {\hspace{1cm}}$0.0 < \sin^{2}2\theta_{13} < 0.15$  & $-$20.6 & $-$17.9 ~~\\
      \hline 
      Total:                  & $+$10.7  & $+$12.0 ~~\\
                              & $-$22.9  & $-$20.3 ~~\\
      \hline 
      \hline
    \end{tabular}
    \caption{Summary of systematic uncertainties for the expected number of
\nutau events (top) and for the observed number of \nutau events (bottom).
The best fit values for each error term are listed for 
both likelihood (LH) and neural network (NN) analyses.} 
    \label{table:systematics}
\end{table}
The systematic errors from \superk atmospheric neutrino oscillation analysis
are re-evaluated for the present analysis;
however in this estimation, the uncertainty in the absolute normalization 
is assumed to be 20\%.
All error terms except for those affecting Sub-GeV, PC, and upward-going 
muon events are considered.
A detailed description of these uncertainties can be found 
in Ref.~\cite{Ashie:2005ik}.
In addition, the uncertainties related only to the present analysis  
such as the \nutau cross section, \nutau polarization, 
tau likelihood, etc. are considered.
The systematic uncertainties for the expected number of 
\nutau events are summarized in Table~\ref{table:systematics}. 

In determining the systematic uncertainties for the observed number of 
\nutau events, various effects (such as up/down ratio) 
that could change the up-down asymmetry of 
the background MC and the data are considered.
The systematic errors due to uncertainties in the oscillation parameters, 
$\Delta m^2_{23}$ and $\sin^{2}2\theta_{23}$, are also estimated 
by using 68\% C.L. allowed parameter region obtained 
by the L/E analysis from \superk~\cite{Ashie:2004mr}. The
uncertainty due to knowledge of the $\sin^{2}2\theta_{13}$ is
estimated using the limit obtained by the CHOOZ reactor neutrino
experiment~\cite{Apollonio:2002gd}.  
This systematic error is asymmetric 
because Multi-GeV electrons are expected to appear 
in the upward-going directions for non-zero \(\theta_{13}\), 
which would be backgrounds for \nutau signals. 
Table~\ref{table:systematics} shows
the summary of systematic uncertainties. We also performed a study to 
check dependency on Monte Carlo neutrino interaction models 
using another model, NUANCE~\cite{Casper:2002sd}.
The difference in the results is negligible.

Combining these errors with the fit result, we obtain a best fit tau neutrino
appearance signal of 138\,$\pm$\,48\,(stat.)\,$^{+15}_{-32}$\,(sys.) from 
the likelihood analysis, which
disfavors the no tau neutrino appearance hypothesis by 2.4 sigma. This is
consistent with the expected number of tau neutrino events, 
78\,$\pm$\,26\,(sys.)
for \dms = $2.4 \times 10^{-3}$\,eV$^2$, assuming the full mixing in
\mutau oscillations.

In conclusion, the search for the appearance of tau neutrinos 
from \mutau oscillations in the atmospheric neutrinos
has been carried out using atmospheric neutrino data observed in  
Super-Kamiokande-I. The tau neutrino excess events have been observed 
in the upward-going direction as expected.
The Super-Kamiokande-I atmospheric neutrino data are consistent with 
\mutau oscillations. 

We gratefully acknowledge the cooperation of the Kamioka Mining and
Smelting Company.  The Super-Kamiokande experiment has been built and
operated from funding by the Japanese Ministry of Education, Culture,
Sports, Science and Technology, the United States Department of Energy,
and the U.S. National Science Foundation.



\end{document}